\documentclass[10pt,twocolumn,letterpaper]{article}

\usepackage[final]{cvpr}      

\usepackage{graphicx}
\usepackage{amsmath}
\usepackage{xcolor}
\usepackage{bbding}
\usepackage{amssymb}
\usepackage{dsfont}
\usepackage{booktabs}
\usepackage{multirow}
\usepackage{makecell}
\usepackage{xcolor}
\usepackage{amsthm}
\usepackage{ cite }
\usepackage{float}
\usepackage{soul}
\usepackage{caption}
\usepackage{subcaption}
\usepackage{graphicx}
\usepackage{thmtools, thm-restate}

\usepackage[normalem]{ulem}
\usepackage[pagebackref,breaklinks,colorlinks]{hyperref}

\usepackage[capitalize]{cleveref}
\crefname{section}{Sec.}{Secs.}
\Crefname{section}{Section}{Sections}
\Crefname{table}{Table}{Tables}
\crefname{table}{Tab.}{Tabs.}

\def\cvprPaperID{5503} 
\def\confName{CVPR}
\def\confYear{2022}

\newcommand\samethanks[1][\value{footnote}]{\footnotemark[#1]}

\begin{document}
\newcommand\ChangeRT[1]{\noalign{\hrule height #1}}

\title{Self-supervised Pseudo Multi-class Pre-training for \\ Unsupervised Anomaly Detection and Segmentation in Medical Images}

\author{
\parbox{.9\linewidth}{\centering $\quad$ Yu Tian$^{1}\thanks{First two authors contributed equally to this work.}$ $\quad$  Fengbei Liu${^{2}\samethanks}$ $\quad$ Guansong Pang$^{5}$ $\quad$  Yuanhong Chen$^{2}$ $\newline$ Yuyuan Liu$^{2}$  $\quad$ Johan W Verjans$^{2,3,4}$ $\quad$ Rajvinder Singh$^{3}$ $\quad$ Gustavo Carneiro$^{6}$ $\newline$  $^{1}$ 
Harvard Ophthalmology AI Lab, Harvard Medical School
\\
 $^{2}$ 
 Australian Institute for Machine Learning, University of Adelaide \\
 $^{3}$ 
 Faculty of Health and Medical Sciences, University of Adelaide
 \\
  $^{4}$ South Australian Health and Medical Research Institute \\
  $^{5}$ School of Computing and Information Systems, Singapore Management University \\
  $^{6}$ Centre for Vision, Speech and Signal Processing, University of Surrey
}
}


\maketitle

\begin{abstract}
Unsupervised anomaly detection (UAD) methods are trained with  normal (or healthy) images only, but during testing, they are able to classify normal and abnormal (or disease) images. 
UAD is an important medical image analysis (MIA) method to be applied in disease screening problems because the training sets available for those problems usually contain only normal images.
However, the exclusive reliance on normal images may result in the learning of ineffective low-dimensional image representations that are not sensitive enough to detect and segment unseen abnormal lesions of varying size, appearance, and shape. 
Pre-training UAD methods with self-supervised learning, based on computer vision techniques, can mitigate this challenge, but they are sub-optimal because they do not explore domain knowledge for designing the pretext tasks, and their contrastive learning losses do not try to cluster the normal training images, which may result in a sparse distribution of normal images that is ineffective for anomaly detection. 
In this paper, we propose a new self-supervised pre-training method for MIA UAD applications, named \underline{P}seudo \underline{M}ulti-class \underline{S}trong \underline{A}ugmentation via \underline{C}ontrastive \underline{L}earning (PMSACL). PMSACL consists of a novel optimisation method that contrasts a normal image class from multiple pseudo classes of synthesised abnormal images, with each class enforced to form a dense cluster in the feature space.
In the experiments, we show that our PMSACL pre-training improves the accuracy of SOTA UAD methods on many MIA benchmarks using colonoscopy, fundus screening and Covid-19 Chest X-ray datasets.  The code is made publicly available via this \href{https://github.com/tianyu0207/PMSACL}{\underline{\textbf{link}}}. 
\end{abstract}

\section{Introduction and Background}
\label{sec:intro}

Detecting and segmenting abnormal lesions from disease screening datasets is a crucial task in medical images analysis (MIA)~\cite{tian2019one,tian2020few,liuphotoshopping,litjens2017survey,baur2020scale,fan2020pranet,lz2020computer,liu2021self,liu2021noisy,liu2022translation,tian2022contrastive,chen2023braixdet,shi2023artifact,chen2022multi,liu2022nvum,wang2022knowledge,chen2022bomd}.

A challenging aspect of this problem is that such screening datasets~\cite{tian2021detecting,pu2019prospective} usually contain a disproportionately large number of normal (or healthy) images, and a tiny amount of abnormal (or disease) images that poorly represent all possible disease sub-classes.
Instead of designing a fully supervised training approach to handle such a heavily imbalanced labelled dataset with a poor representation of disease sub-classes, we consider in this paper an alternative approach based on unsupervised anomaly detection (UAD)~\cite{tian2021constrained,chen2021unsupervised,chen2020unsupervised}, which is trained exclusively with normal images.
There are two advantages with such UAD strategy: 1) the acquisition of the training set is straightforward given the large proportion of normal images in screening datasets; and 2) it is not necessary to collect a representative training set containing abnormal images from all disease sub-classes.
Nevertheless, this UAD strategy is challenging because the model needs to classify abnormal images without being exposed to them during training.

UAD methods are generally based on a one-class classifier (OCC) that learns a normal image distribution from the normal training images, and test image anomalies (or abnormal images) are detected based on the extent that they deviate from the learned distribution~\cite{f-AnoGAN,seebock2019exploiting,gong2019memorizing,chen2021unsupervised,liu2019photoshopping,venkataramanan2020attention,tian2021weakly,chen2020unsupervised,tian2020few,pang2019deep,reiss2021mean,tian2022pixel,tian2022unsupervised}.
One fundamental problem in such UAD methods is the learning of effective image feature representations.
To detect lesions in medical images, this problem is particularly critical because anomalous lesions may be represented by subtle deviations from normal tissues (e.g., tiny and flat colon polyps)~\cite{tian2020few}. 
If not well trained, these UAD methods can overfit the normal training data and learn ineffective image representations, failing to detect and segment subtle abnormal lesions.
Previous papers tackle this problem with the use of ImageNet-based pre-trained models, but transferring representations learned from natural images to medical images may not be optimal and can deteriorate the detection performance in medical domain~\cite{tian2021constrained}.

Another pre-training approach is based on self-supervised learning (SSL)~\cite{hendrycks2019using,golan2018deep,bergman2020classification,simclr_paper,moco,liu2020self,tian2021constrained,cho2021self,self-medmix-new-ref,self-medmix-new-ref2}, whose effectiveness depends on the relatedness of the pretext tasks and the final MIA classification task, and on assumptions about the training process.
SSL pre-training for UAD methods applied to MIA screening problems have shown promising results~\cite{tian2021constrained}, but they have been sub-optimally explored given that they were adapted from computer vision methods without using MIA domain knowledge in the design of the pretext tasks or in the training process.
For instance, even though normal samples can potentially form multiple sub-clusters in the representational space due to the appearance variability within normal training samples, the particular number of such sub-clusters is unknown and hard to define.
Hence, in MIA, normal images can be assumed to form a single class, while disease images can be divided into sub-classes characterised by variations in the number and appearance of lesions. 
Nevertheless, previous SSL methods in UAD~\cite{tack2020csi,tian2021constrained,sohn2020learning} extend contrastive learning~\cite{simclr_paper} by sub-dividing the normal class images into multiple classes charaterised by geometric or appearance transformations.
Unfortunately, such training process 
is challenging for MIA UAD 
that needs to discriminate a single dense class of normal images against a relatively small number of abnormal sub-classes that lie outside the normal class distribution.

In this paper, we propose the \underline{P}seudo \underline{M}ulti-class \underline{S}trong \underline{A}ugmentation via \underline{C}ontrastive \underline{L}earning (PMSACL), a new self-supervised pre-training method modelled exclusively with normal training images, and designed to learn effective image representations for different types of downstream UAD methods applied to several MIA problems. 
The main advantage of PMSACL, compared to previous self-supervised pre-training method for MIA applications~\cite{tian2021constrained}, is that we rely on MIA domain knowledge to design the training and the pretext tasks. 
In particular, our training uses contrastive learning to classify training samples into multiple dense clusters in terms of Euclidean distance and cosine similarity, with one cluster  representing the normal images and the remaining ones representing pseudo sub-classes of the disease images.
These pseudo disease sub-classes are synthesised with our MedMix augmentations that simulate a varying number of lesions of different sizes and appearance in the normal training images (see Fig.~\ref{fig:MedMix-example}). Please note that our proposed MedMix is inspired by the previous methods based on cut-paste/cut-mix operations from natural and medical images~\cite{cho2021self,miccai21challenge,self-medmix-new-ref,self-medmix-new-ref2}. 
We summarise our contributions as follows:
\begin{itemize}
\item Our PMSACL is the first self-supervised pre-training method specifically designed for MIA UAD applications, where the main advantage lies in the contrastive learning optimisation that learns multiple classes, one for normal images, and the others for pseudo sub-classes of disease images, which are synthesised by our MedMix augmentations by simulating a varying number of lesions of different sizes and appearance.
\item We extend our previously published CCD method~\cite{tian2021constrained} by proposing two new loss functions to form denser clusters per class, namely: 1) a multi-centring loss to constrain the feature representations of different classes into a subspace around their class centres; and 2) a non-trivial extension of the normalisation of the standard contrastive loss that repels samples from the same class with less intensity than the samples from different classes.
\item The proposed PMSACL is shown to learn effective image representations that can adapt well to different types of downstream UAD methods applied to several MIA problems. 
\end{itemize}
We empirically show that PMSACL pre-training significantly improves the performance of two SOTA anomaly detectors, PaDiM~\cite{defard2020PaDiM} and IGD~\cite{chen2021unsupervised}. 
Extensive experimental results on four different disease screening medical imaging benchmarks, namely, colonoscopy images from two datasets~\cite{borgli2020hyperkvasir,liu2019photoshopping}, fundus images for glaucoma detection~\cite{li2019attention} and Covid-19 Chest X-ray (CXR) dataset~\cite{wang2020covid} show that PMSACL can be used to pre-train diverse SOTA UAD methods to improve their accuracy in detecting and segmenting lesions in diverse medical images.

\textbf{Relationship to Preliminary Work:} An early version of this work was presented in our previously published paper~\cite{tian2021constrained}. In this new submission, we considerably expand that previous study by: 1) resolving the strong augmentations issue of CCD that does not synthesise medical image anomalies that are relevant for downstream UAD applications;
2) addressing the issues around CCD's contrastive learning that does not consider that the downstream UAD methods will classify one class of normal images and a
few sub-classes of disease images; 
3) providing a more comprehensive literature review; 4) adding more experiments using datasets from many medical domains; and 5) including a more in-depth analysis of the proposed PMSACL.

\section{Related Work}


\textbf{UAD approaches}~\cite{f-AnoGAN,seebock2019exploiting,gong2019memorizing,chen2021unsupervised,liu2019photoshopping,venkataramanan2020attention,tian2021weakly,chen2020unsupervised,tian2020few,pang2019deep,liu2022residual} can be divided into two categories: predictive-based (e.g., DSVDD~\cite{ruff2018deep}, OC-SVM~\cite{chen2001one}, and deviation network~\cite{pang2019deep}), and generative-based (e.g., auto-encoder~\cite{venkataramanan2020attention,chen2020unsupervised,chen2021unsupervised,gong2019memorizing} and GAN~\cite{f-AnoGAN,liuphotoshopping,akcay2018ganomaly}). 
Predictive-based UAD approaches train a one-class classifier to describe the distribution of normal data, and discriminate abnormal data using their distance/deviation to the normal data distribution; whereas generative-based UAD approaches train deep generative models to learn latent representations of normal images, and detect anomalies based on image reconstruction error~\cite{pang2021deep}. 
A fundamental challenge in both types of UAD methods is the learning of expressive feature representations from images, which is particularly important in MIA because abnormal medical images may have subtly looking lesions that can be hard to differentiate from normal images.
Hence, if not well trained, these UAD models can become over-confident in classifying normal training data and learn ineffective image representations that will fail to enable the detection and segmentation of lesions.

\textbf{Pre-training} is an effective approach to address the representation challenge described above. A heavily explored pre-training approach is based on using ImageNet~\cite{deng2009imagenet} pre-trained models, but transferring representations learned from natural images to medical images is not straightforward~\cite{tian2021constrained}. 
Alternatively, the representation challenge can also be tackled by pre-training methods based on self-supervised learning (SSL) that learns auxiliary pretext tasks~\cite{hendrycks2019using,golan2018deep,bergman2020classification,simclr_paper,moco,liu2020self}. SSL is a strategy that has produced effective representations for UAD in general computer vision tasks~\cite{hendrycks2019using,golan2018deep,bergman2020classification,tack2020csi}.
However, their application to MIA problems needs to be further investigated because it is not clear how to design effective training or pretext tasks that can work well in the detection of subtle lesions in medical images.
Previous UAD methods relied on self-supervised pretext tasks based on the prediction of geometric transformations~\cite{hendrycks2019using,golan2018deep,bergman2020classification} or contrastive learning using standard data augmentation techniques (e.g., scaling, cropping, etc.)~\cite{simclr_paper,moco} to form a large number of image classes characterising similar and dissimilar pairs.
These pretext tasks and training strategy are not specifically related to the detection of subtle anomalies in medical images that contain a normal image class and a small number of disease sub-classes, so they may even degrade the detection accuracy of downstream UAD methods~\cite{wang2020understanding}. 

For SSL UAD pre-training in MIA, the only previous work that we are aware is our previously published CCD method~\cite{tian2021constrained} that adapts standard contrastive learning and two general computer vision pretext tasks to image anomaly detection and can be applied to multiple downstream UAD methods. Although achieving good results in many benchmarks,
the training explored by CCD does not explore the fact that the downstream UAD methods need to recognise one class of normal images and a small number of sub-classes of disease images, and the CCD's data augmentation will not 
synthesise relevant medical image anomalies -- both issues can challenge the training of  downstream UAD approaches. 


\section{Method}

\begin{figure}
\begin{center}
\includegraphics[width=1.0 \linewidth]{ 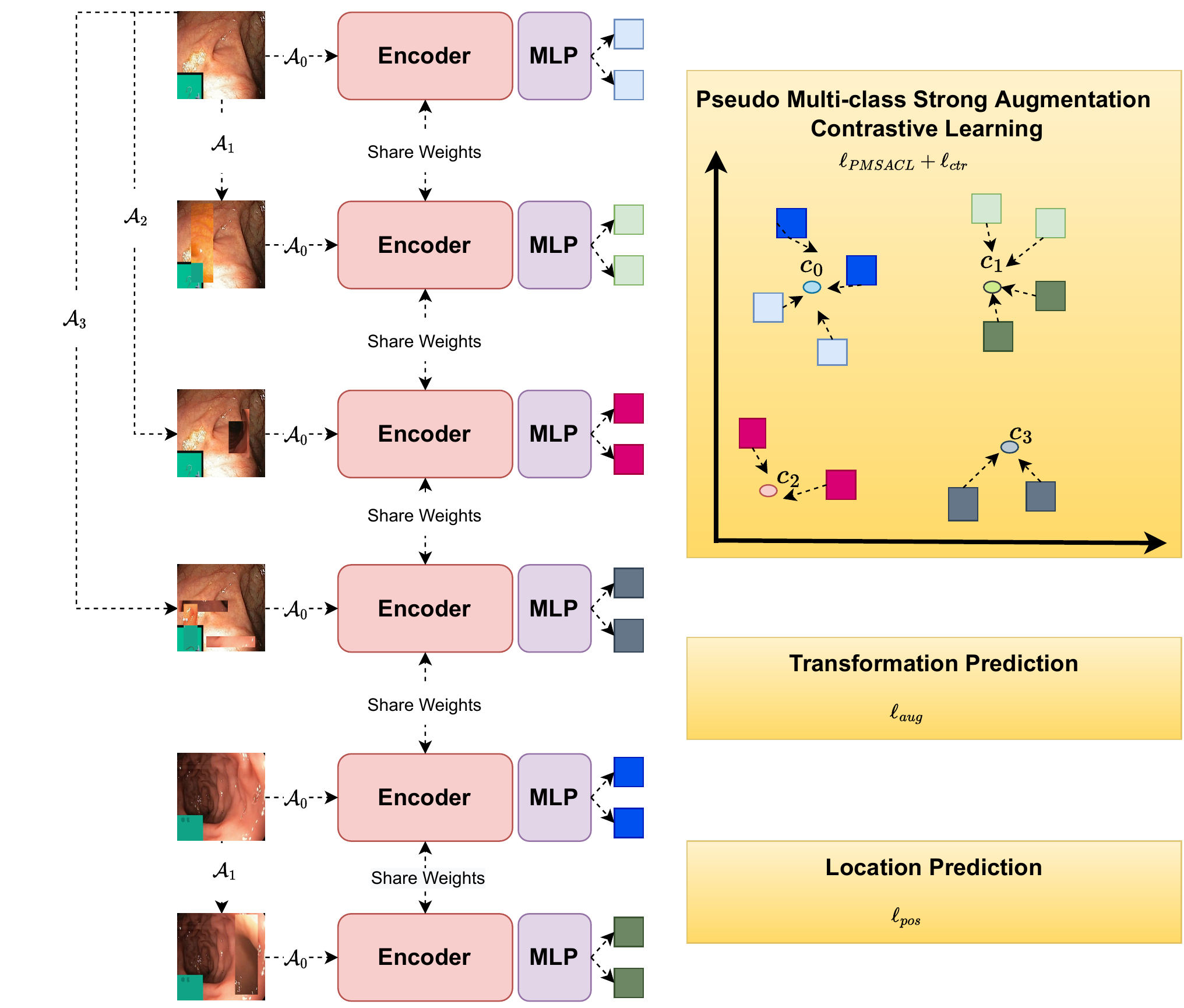}
\end{center}
   \caption{\textbf{PMSACL}: our proposed self-supervised pre-training for UAD trains four classes of images: the normal images formed by the weak augmentations in distribution $\mathcal{A}_0$ (blue markers) and three classes of synthetised abnormal images formed by the strong augmentation in distributions $\{\mathcal{A}_n\}_{n=1}^{3}$ (green, pink and orange markers). The optimisation uses a constrained contrastive learning that trains a four-class classification problem.
   The different types of strong augmentations are produced by MedMix that introduces a varying number of fake lesions by cutting patches from the normal training images, altering them with random color jittering, Gaussian noise and non-linear intensity transformations, and pasting them to other normal training images.}
\label{fig:framework}
\end{figure}

In this section, we introduce the proposed PMSACL pre-training approach depicted in Fig.~\ref{fig:framework}. Given a training medical image dataset $\mathcal{D} = \{ \mathbf{x}_i \}_{i=1}^{|\mathcal{D}|}$, 
with all images assumed to be from the $\text{normal}$ class and $\mathbf{x} \in \mathcal{X} \subset \mathbb{R}^{H \times W \times C}$ ($H$: height, $W$: width, $C$: number of colour channels), our learning strategy involves two stages: 1) the self-supervised pre-training to learn an encoding network $f_{\theta}:\mathcal{X} \rightarrow \mathcal{Z}$ (with $\mathcal{Z} \subset \mathbb{R}^{Z}$), and
2) the fine-tuning of an anomaly detector or segmentation model 
built from the pre-trained $f_{\theta}(.)$.  
The approach is evaluated on a testing set $\mathcal{T} = \{ (\mathbf{x},y,\mathbf{m})_i \}_{i=1}^{|\mathcal{T}|}$, where $y \in \mathcal{Y} = \{\text{normal}, \text{abnormal} \}$, and $\mathbf{m}\in \mathcal{M} \subset \{0,1\}^{H \times W \times 1}$ denotes the segmentation mask of the lesion in the image $\mathbf{x}$. Below, we first introduce the MedMixdata augmentation in Sec.~\ref{sec:medmix}, then describe the optimisation proposed for PMSACL in Sec.~\ref{sec:MSACL}, followed by a bried description of the UAD methods in Sec.~\ref{sec:anomaly_detection_Segmentation}.


\subsection{MedMix Augmentation}
\label{sec:medmix}
Our MedMix augmentation is designed to augment medical images to simulate multiple lesions. 
We target a more effective data augmentation for MIA applications than the computer vision  augmentations in~\cite{tian2021constrained} (e.g., permutations, rotations) that do not simulate medical image anomalies and may yield poor detection performance by downstream UAD methods.
We realise that anomalies in different medical domains (e.g., glaucoma and colon polyps) can be visually different, but a commonality among anomalies is that they are usually represented by an unusual growth of abnormal tissue. 
Hence, we propose the MedMix augmentation to simulate abnormal tissue with a strong augmentation that ``constructs" abnormal lesions by the cutting and pasting (from and to normal images) of small and visually deformed patches. 
This visual deformation is achieved by applying other transformations to patches, such as colour jittering, Gaussian noise and non-linear intensity transformations.
This approach is inspired by cutmix~\cite{yun2019cutmix} and CutPaste~\cite{li2021cutpaste}, where our contribution over those approaches is the intensification of the change present in the cropped patches by the appearance transformations above. 
These transformations are designed to encourage the model to learn abnormalities in terms of localised image appearance, structure, texture, and colour.

\begin{figure}[t]
\begin{center}
\includegraphics[width=0.95\linewidth]{ 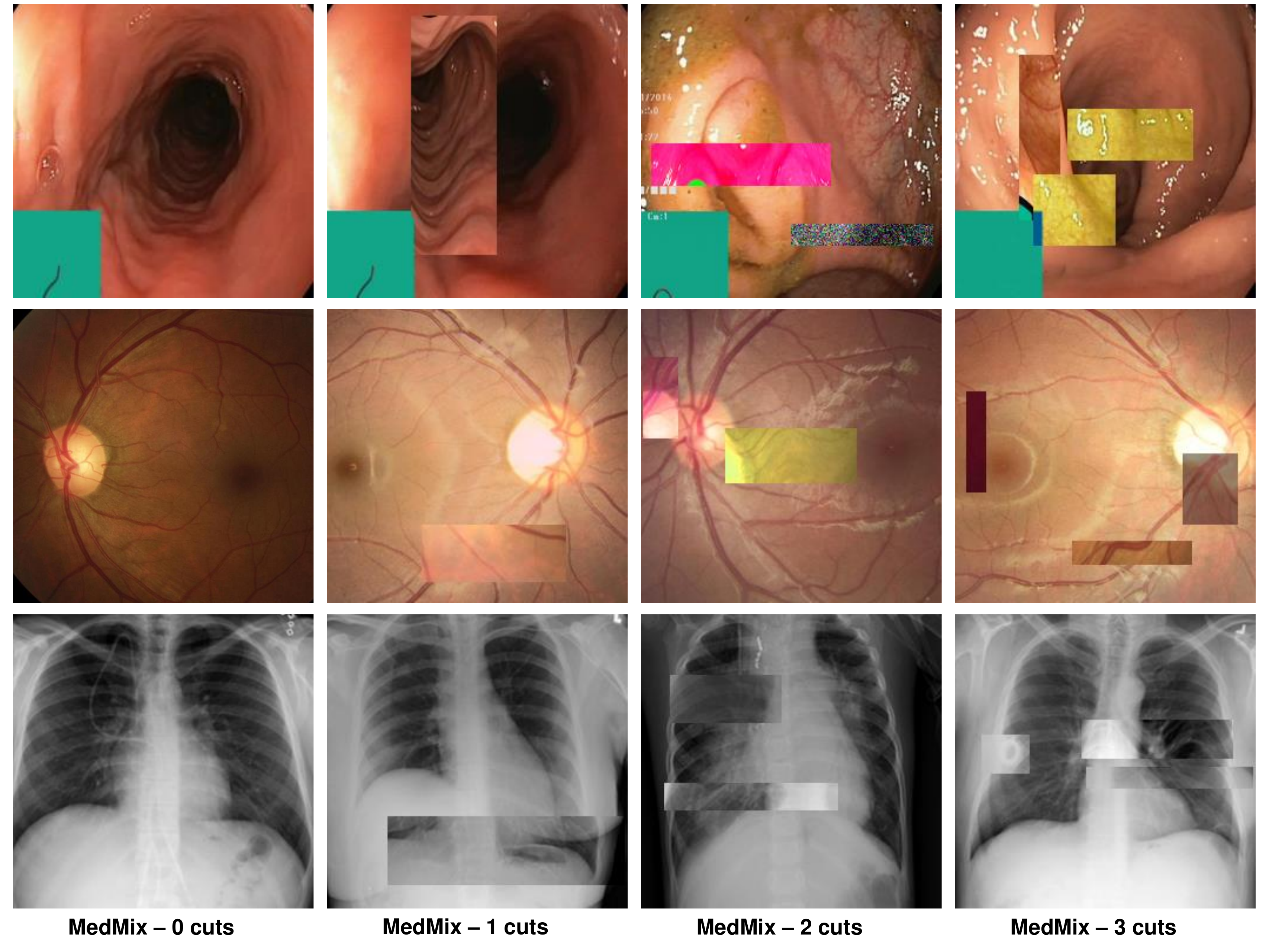}
\end{center}
   \caption{Examples of our MedMix data augmentation, showing  augmentation $\mathcal{A}_0$ containing zero synthetic anomalies (leftmost column) and increasingly stronger augmentations $\{\mathcal{A}_n\}_{n=1}^3$ (second to fourth columns) with different number of synthetic anomalies (from one to three).}
\label{fig:MedMix-example}
\end{figure}

In practice, we design $|\mathcal{A}|=4$ strong augmentation distributions, where 
$\mathcal{A}_n$ includes $n\in\{0,...,3\}$ abnormalities in the image, which means that $\mathcal{A}_0$ denotes the normal image distribution and $\mathcal{A}_{n\in\{1,2,3\}}$ represent the abnormal image distributions, containing $\{1,2,3\}$ anomalous regions.
Therefore, our loss targets the classification of MedMix augmentations, as shown in Fig.~\ref{fig:MedMix-example}. 

\subsection{PMSACL Pre-training}
\label{sec:MSACL}

The gist of our proposed PMSACL lies in the idea of discriminating the distribution of weakly augmented samples (simulating normal images) from the distributions of different types of strongly augmented samples (simulating multiple classes of abnormal images).
Instead of attracting and repelling samples within and between a large number of image classes~\cite{tack2020csi,tian2021constrained,sohn2020learning}, we propose a new contrastive loss to separate samples from the normal class and samples from pseudo abnormal sub-classes, and to enforce the clusters representing the normal and abnormal sub-classes to be dense. 
To this end, our proposed loss is defined as: 
\begin{equation}
\resizebox{0.96\linewidth}{!}{%
$
\begin{split}
\ell(\mathcal{D};\theta,\beta,\gamma)= 
    \ell_{ctr}(\mathcal{D};\theta) + \ell_{PMSACL}(\mathcal{D};\theta) + 
    \ell_{aug}(\mathcal{D};\beta) + \ell_{pos}(\mathcal{D};\gamma), 
\end{split}
$}
\label{eq:full_loss}
\end{equation}
where $\ell_{ctr}(.)$ denotes the new distribution multi-centring loss, $\ell_{PMSACL}(.)$ represents the new PMSACL contrastive loss, $\ell_{aug}(.)$ and $\ell_{pos}(.)$ are the pretext learning losses to regularise the optimisation~\cite{tian2021constrained}, and $\theta$, $\beta$ and $\gamma$ are trainable parameters. 
The loss terms in~\eqref{eq:full_loss} rely on \textbf{weak} data augmentation distribution, denoted by $\mathcal{A}_0$, and $|\mathcal{A}|$ \textbf{strong} data augmentation distributions, represented by $\{ \mathcal{A}_n \}_{n=1}^{|\mathcal{A}|}$, each denoting a different type of augmentation. From each of these distributions, we can sample augmentation functions $a:\mathcal{X} \to \mathcal{X}$. 

The multi-centring loss in~\eqref{eq:full_loss} depends on the estimation of the mean representation for each augmentation distribution, computed as
\begin{equation}
\mathbf{c}_n = \mathbb{E}_{\mathbf{x} \in \mathcal{D},  a \sim \mathcal{A}_n}  [ f_{\theta}(a(\mathbf{x})) ],
\label{eq:compute_center}
\end{equation}
where $n \in \{0,...,|\mathcal{A}|\}$, with $\mathbf{c}_n$ being the mean representation of the training data augmented by the functions sampled from $\mathcal{A}_n$. Note that these mean representations are computed at the beginning of the training and frozen for the rest of the training.
The distribution multi-centring loss is then defined as: 
\begin{equation}
\ell_{ctr}(\mathcal{D};\theta) = \mathbb{E}_{\mathbf{x} \in \mathcal{D},  n \in \{0,...,|\mathcal{A}|\}, a \sim \mathcal{A}_n  }   \| f_{\theta}(a(\mathbf{x}))  - \mathbf{c}_n  \|^2,
\label{eq:center_loss}
\end{equation}
which pulls the representations of augmented samples toward their mean representations in~\eqref{eq:compute_center}, making the augmentation clusters dense in Euclidean space. 

To further enforce the separation between different clusters and the tightness within each cluster, we introduce a novel contrastive learning loss function. 
In our contrastive learning, we maximisise the cosine similarity of samples that belong to the same class (i.e., normal or one of the abnormal sub-classes) and minimise the cosine similarity of samples belonging to different classes.
An interesting aspect of this optimisation is that samples are centred by their own cluster mean representation $\mathbf{c}_n$ from~\eqref{eq:compute_center}, so our contrastive learning, combined with the multi-centred loss in~\eqref{eq:center_loss} will cluster samples of the same class not only in Euclidean space, but also in inner product space (with cosine measuring similarity between samples).
Such re-formulated constrastive learning, combined with the multi-centring loss~\eqref{eq:center_loss}, results in a loss that produces multiple clusters, where cluster $n=0$ contains the normal images and the others, denoted by $n\in\{1,...,|\mathcal{A}|\}$), have the synthetised abnormal images.
Our PMSACL loss is defined as:
\begin{equation}
\resizebox{0.96\linewidth}{!}{%
$
\begin{split}
    \ell_{PMSACL}(\mathcal{D};\theta)=\mathbb{E}_{\mathbf{x} \in \mathcal{D},n\in\{0,...,|\mathcal{A}|\},l \in \{0,1\} } \left [ \ell^x_{PMSACL}( \mathbf{x}^{(n,l)}, \mathcal{D};\theta) \right ]
\end{split}
$}
 \label{eq:contrastive_loss}
\end{equation}
where $\mathbf{x}^{(n,l)} = a(\mathbf{x}^{(n)})$
represents one of two (indexed by $l\in\{0,1\}$) augmented data obtained from the application of a weak augmentation $a \sim \mathcal{A}_0$ on a strongly augmented data denoted by $\mathbf{x}^{(n)} = a(\mathbf{x})$ with $a \sim \mathcal{A}_n$. 
In~\eqref{eq:contrastive_loss}, we have:
\begin{equation}
\resizebox{0.96\linewidth}{!}{
$
\begin{split}
    \ell^x_{PMSACL}( \mathbf{x}^{(n,l)}, \mathcal{D};\theta)= 
    -\log\frac{\exp\left[ \frac{1}{\tau}  (\mathbf{f}^{(n,l)})^{\top} 
      \mathbf{f}^{(n,(l+1)\,\text{mod}\,{2})}\right]}
     {
    \sum\limits_{\substack{\mathbf{x}_j \in \mathcal{D} \\ m \in \{0,...,|\mathcal{A}|\} \\ k \in \{0,1\}}}
     \mathbb{I}(\mathbf{x}_j^{(m,k)} \neq \mathbf{x}^{(n,l)}) \exp\left[ \kappa(n, m)   (\mathbf{f}^{(n,l)})^{\top}  \mathbf{f}^{(m,k)}) \right]
     },
\end{split}
$}
\label{eq:sample_constrastive_loss}
\end{equation}
where $\mathbb{I}(.)$ denotes an indicator function,  $\mathbf{x}_j^{(m,k)}$ is defined similarly as $\mathbf{x}^{(n,l)}$ in~\eqref{eq:contrastive_loss},
$m \in \{0,...,|\mathcal{A}|\}$ indexes the set of strong augmentations, and $k \in \{0,1\}$ indexes one of the two weak augmentations applied to the strongly augmented image. 
Lastly, to further constrain the normal and strongly augmented data representations in~\eqref{eq:contrastive_loss}, our PMSACL loss minimises the distance between samples centred by their representation means computed as:
\begin{equation}
 \mathbf{f}^{(n,l)} = \frac{f_{\theta}(\mathbf{x}^{(n,l)}) - \mathbf{c}_n}{\| f_{\theta}(\mathbf{x}^{(n,l)}) - \mathbf{c}_n \|_2} , 
    \label{eq:temperature_calibration}
\end{equation}
where $\mathbf{c}_n$ is defined in~\eqref{eq:compute_center}.
Also in~\eqref{eq:contrastive_loss} to map the representations from the same distribution into a denser region of the hyper-sphere~\cite{cho2021masked}, we propose a temperature scaling strategy defined as: 
\begin{equation}
\kappa(n,m)= 
\begin{cases}
1 /(\alpha\tau)   &, \text{if } n = m \\

1 /\tau              &, \text{otherwise}
\end{cases},
    \label{eq:temperature_calibration}
\end{equation}
where $\alpha$ is a scaling factor that controls the shrinkage level of the temperature $\tau$.
As a result, Eq.~\eqref{eq:temperature_calibration} alters the temperature for the samples that belong to the same strong augmentation distributions (i.e., when $n = m$) to a smaller value $\alpha$, which allows smaller amount of repelling strength compared to samples that belongs to strong augmentation distributions (i.e., $n \neq m$).
Putting all together, the loss in~\eqref{eq:contrastive_loss} clusters the image representations into  hyper-spheres and regions within the hyper-spheres, where each hyper-sphere and region represent a different type of augmentation. 

Inspired by~\cite{tack2020csi,tian2021constrained}, we further constrain the training in~\eqref{eq:full_loss} with a self-supervised classification constraint $\ell_{aug}(\cdot)$ that enforces the model to classify the strong augmentation function (Fig.~\ref{fig:framework}):
\begin{equation}
\ell_{aug}(\mathcal{D};\beta) = -\mathbb{E}_{\mathbf{x} \in \mathcal{D}, n\in\{0,...,|\mathcal{A}|\}, a \sim \mathcal{A}_n } \left [ \log \mathbf{a}_n^{\top} f_{\beta}(f_{\theta}(a(\mathbf{x}))) \right ],
\label{eq:pretext_loss}
\end{equation}
where $f_{\beta}:\mathcal{Z} \rightarrow [0,1]^{|\mathcal{A}|}$ is a fully-connected (FC) layer, and $\mathbf{a}_n \in \{0,1\}^{|\mathcal{A}|}$ is a one-hot vector representing the strong augmentation distribution (i.e., $\mathbf{a}_n(j)=1$ for $j=n$, and $\mathbf{a}_n(j)=0$ for $j \neq n$).

The final constraint in~\eqref{eq:full_loss} is based on the relative patch location from the centre of the training image and is adapted for local patches. 
This constraint is added to learn  positional and texture characteristics of the image in a self-supervised manner. 
Inspired by~\cite{doersch2015unsupervised}, the positional constraint predicts the relative position of the paired image patches, with its loss defined as
\begin{equation}
\ell_{pos}(\mathcal{D};\gamma) = -\mathbb{E}_{ \{ \mathbf{x}_{\omega_{1}},\mathbf{x}_{\omega_{2}} \} \sim \mathbf{x} \in \mathcal{D}} \left [ \log \mathbf{p}^{\top} f_{\gamma}(f_{\theta}(\mathbf{x}_{\omega_{1}}),f_{\theta}(\mathbf{x}_{\omega_{2}})) \right ],
\label{eq:patch_prediction}
\end{equation}
where $\mathbf{x}_{\omega_{1}}$ is a randomly selected fixed-size image patch from $\mathbf{x}$, $\mathbf{x}_{\omega_{2}}$ is another image patch from one of its eight neighbouring patches, 
$\omega_1,\omega_2 \in \Omega$ represents indices to the image lattice,
$f_{\gamma}:\mathcal{Z}\times\mathcal{Z} \rightarrow [0,1]^8$, 
and $\mathbf{p} = \{0, 1\}^8$ is a one-hot encoding of the patch location.
The constraints in~\eqref{eq:pretext_loss} and~\eqref{eq:patch_prediction} are designed to improve training regularisation.


\subsection{Anomaly Detection and Segmentation}
\label{sec:anomaly_detection_Segmentation}


After pre-training $f_\theta(\cdot)$ with PMSACL, we fine-tune it with a SOTA UAD, such as IGD~\cite{chen2021unsupervised} or PaDiM~\cite{defard2020PaDiM}.
Those methods use the same training set $\mathcal{D}$ as PMSACL, containing only normal images from healthy patients.

IGD~\cite{chen2021unsupervised} combines three loss functions: 1) two reconstruction losses based on local and global multi-scale structural similarity index measure (MS-SSIM)~\cite{wang2003multiscale} and mean absolute error (MAE) to train the encoder $f_{\theta}:\mathcal{X} \to \mathcal{Z}$ and decoder $g_{\phi}:\mathcal{Z} \to \mathcal{X}$, 2) a regularisation loss to train adversarial interpolations from the encoder~\cite{berthelot2018understanding}, and 3) an anomaly classification loss to train $h_{\psi}:\mathcal{Z} \to [0,1]$. The anomaly detection score of image $\mathbf{x}$ is defined by
\begin{equation}
    s_{IGD}(\mathbf{x}) = \xi \ell_{rec}(\mathbf{x},\tilde{\mathbf{x}}) + (1-\xi)(1 - h_{\psi}(f_{\theta}(\mathbf{x}))),
    \label{eq:anomaly_score_image}
\end{equation}
where $\tilde{\mathbf{x}} =g_{\phi}(f_{\theta}(\mathbf{x}))$,
$h_{\psi}(.)$ returns the likelihood that $\mathbf{x}$ is a normal image, $\xi \in [0,1]$ is a hyper-parameter, and
\begin{equation}
\resizebox{0.96\linewidth}{!}{%
$
\begin{split}
    \ell_{rec}(\mathbf{x},\tilde{\mathbf{x}}) = 
    \rho \| \mathbf{x} - \tilde{\mathbf{x}} \|_1 + 
     (1-\rho) \left ( 1-\left( 
     \nu m_{G}(\mathbf{x},\tilde{\mathbf{x}}) + (1-\nu)m_{L}(\mathbf{x},\tilde{\mathbf{x}}) \right ) \right ),
\end{split}
$}
    \label{eq:reconstruction error}
\end{equation}
with $\rho,\nu \in [0,1]$, $m_{G}(\cdot)$ and $m_{L}(\cdot)$ denoting the global and local MS-SSIM scores from the global and local models, respectively~\cite{chen2021unsupervised}. 
In particular, the Gaussian Anomaly classifier $h_{\psi}$ constrains the latent space of the IGD encoder with the following loss:
\begin{equation}
    \ell_h(\mathbf{x}) =  h_{\psi}(f_{\theta}(\mathbf{x})),
    \label{eq:loss_GSVDD}
\end{equation}
where the classifier is defined with $h_{\psi}(f_{\theta}(\mathbf{x})) = 1-\exp\left (  - \frac{\| f_{\theta}(\mathbf{x}) - \mu_{\gamma}  \|_2^2}{\sigma_{\gamma}^2} \right )$,  
$f_{\theta}:\mathcal{X} \to \mathcal{Z}$ represents the encoder parameterised by $\theta$, with $\mathcal{Z} \subset \mathbb{R}^Z$ denoting the space of latent embeddings of the auto-encoder.  
The mean and standard deviation values above are estimated with $\mu_{\gamma} = \frac{1}{|\mathcal{D}|}\sum_{i=1}^{|\mathcal{D}|}f_{\theta}(\mathbf{x}_i)$ and $\sigma^2_{\gamma} = \frac{\kappa}{|\mathcal{D}|} \sum_{i=1}^{|\mathcal{D}|} \|f_{\theta}(\mathbf{x}_{i})-\mu_{\gamma}\|_2^2$, 
where $\kappa \in (0,1]$ 
is a constant that regularises the estimation of $\sigma^2_{\gamma}$ to prevent numerical instabilities during training. Such an anomaly classifier optimisation has shown to be less sensitive to outliers compared with other anomaly detection approaches. 
Anomaly segmentation uses~\eqref{eq:anomaly_score_image} to compute $s_{IGD}(\mathbf{x}_{\omega})$, $\forall \omega \in \Omega$ using global and local models, where
$\mathbf{x}_{\omega} \in \mathbb{R}^{\hat{H} \times \hat{W} \times C}$ is an image patch.  This forms a heatmap, where large values of $s_{IGD}(.)$ denote anomalous regions. 
The final heatmap is formed by summing up the global and local heatmaps.

PaDiM~\cite{defard2020PaDiM} utilises the multi-layer features from the pre-trained network $f_{\theta}(.)$ to learn a position dependent multi-variate Gaussian distribution of normal image patches.
Training uses samples collected from the concatenation of the multi-layer features from each patch position $\omega \in \Omega$ to learn the mean and covariance of the Gaussian model denoted by  $\mathcal{N}(\mu_{\omega},\Sigma_{\omega})$~\cite{defard2020PaDiM}.
Anomaly detection is based on the Mahalanobis distance between the concatenated testing patch feature $\mathbf{x}_{\omega}$ and the learned Gaussian distribution $\mathcal{N}(\mu_{\omega},\Sigma_{\omega})$ at that patch position $\omega \in \Omega$ to provide a score of each patch position~\cite{defard2020PaDiM}. 
In particular, anomaly segmentation is inferred using the following anomaly score map:
\begin{equation}
    s_{PaDiM}(\mathbf{x}_{\omega})=  \sqrt{(\mathbf{x}_{\omega} - \mu_{\omega})^{\top}\Sigma_{\omega}^{-1} (\mathbf{x}_{\omega} - \mu_{\omega})},
\end{equation}
and the final score of the whole image $\mathbf{x}$ is defined as:
$s_{PaDiM}(\mathbf{x})=\max_{\omega \in \Omega} s_{PaDiM}(\mathbf{x}_{\omega})$.

\section{Experiments}
\label{sec:experiments}


\subsection{Datasets}
\label{sec:dataset}

We test our self-supervised pre-training PMSACL on four health screening datasets, where we run experiments for both anomaly detection and localisation.
The datasets for anomaly detection and localisation are: the colonoscopy images of Hyper-Kvasir dataset~\cite{borgli2020hyperkvasir}, and the glaucoma dataset using fundus images~\cite{li2019attention}.
We also run anomaly detection without localisation experiments on the following datasets: Liu et al.'s colonoscopy dataset~\cite{liu2019photoshopping}, and Covid-19 chest ray dataset~\cite{wang2020covid} -- these two datasets do not have lesion segmentation annotations, so we test anomaly detection only.

\textbf{Hyper-Kvasir} is a large multi-class public  gastrointestinal imaging dataset~\cite{borgli2020hyperkvasir}. 
We use a subset of the normal (i.e., healthy) images from the dataset for training. 
Specifically, 2,100 images from `cecum', `ileum' and `bbps-2-3' are selected as normal, from which we use 1,600 for training and 500 for testing. We also take 1,000 abnormal images and their segmentation masks of polyps to be used exclusively for testing, where all images have size 300 $\times$ 300 pixels.  

\textbf{LAG} is a large scale fundus image dataset for glaucoma diagnosis~\cite{li2019attention}. For the experiments, we use 2,343 normal (negative glaucoma) images for training, and 800 normal images and 1,711 abnormal images with positive glaucoma with annotated attention maps by ophthalmologists for testing, where images are 500 $\times$ 500 pixels. The annotated attention maps are based on eye tracking, in which the maps are used by the ophthalmologists to explore the region of interest for glaucoma diagnosis~\cite{li2019attention}. 


\textbf{Liu et al.'s colonoscopy dataset} is a colonoscopy image dataset with 18 colonoscopy videos from 15 patients~\cite{liu2019photoshopping}. The training set contains 13,250 normal (healthy) images without polyps, and the testing set contains 967 images, with 290 abnormal images with polyps and 677 normal (healthy) images without polyps, where all images have size  64 $\times$ 64 pixels.

\textbf{Covid-X}~\cite{wang2020covid} has a training set with 1,670 Covid-19 positive and 13,794 Covid-19 negative CXR images. 
The test set contains 400 CXR images, consisting of 200 positive and 200 negative images. We train the methods with the 13,794 Covid-19 negative CXR training images and test on the 400 CXR images, where images are 299 $\times$ 299 pixels.

\subsection{Implementation Details}
\label{sec:implementation_details}

For the proposed PMSACL pre-training, we use Resnet18~\cite{he2016deep} as the backbone architecture for the encoder $f_{\theta}(\mathbf{x})$, and similarly to previous works~\cite{simclr_paper,sohn2020learning}, we add an MLP to this backbone as the projection head for the contrastive learning, which outputs features in $\mathcal{Z}$ of size 128. 
All images from the Hyper-Kvasir~\cite{borgli2020hyperkvasir}, LAG~\cite{li2019attention} and Covid-X~\cite{wang2020covid} datasets are resized to 256 $\times$ 256 pixels. 
For the Liu et al.'s colonoscopy dataset~\cite{liu2019photoshopping}, we use the original image size of 64 $\times$ 64 pixels.  
The batch size is set to 32 and learning rate to 0.01 for the self-supervised pre-training on all datasets.
The model is trained using stochastic gradient descent (SGD) optimiser with momentum. 

We investigate the impact of different strong augmentations in $\mathcal{A}_n$, including rotation, permutation, cutout, Gaussian noise and our proposed MedMix. 
For MedMix patches, we randomly apply colour jittering, Gaussian noise, and  non-linear intensity transformations (i.e., fisheye and horizontal wave transformations). 
In particular, the probability of applying the colour jittering, Gaussian noise, fisheye, and horizontal wave transformations is 25\%. For the patch selection and placement, we randomly cropped a patch from a random image within the minibatch and randomly paste the cropped patch into a random location of another random image from the minibatch. For non-linear transformation, three randomly selected control points are used to alter the MedMix cropped patches and we follow the default settings from the opencv library. 
The weak augmentations in $\mathcal{A}_0$ are the same as in SimCLR~\cite{simclr_paper}, namely: colour jittering, random greyscale, crop, resize, and Gaussian blur. 
We use the same weak augmentation hyper-parameters as the original SimCLR paper (i.e., Gaussian kernel parameters, crop resolution, etc).

The model pre-trained with PMSACL is fine-tuned with  IGD~\cite{chen2021unsupervised} or PaDiM~\cite{defard2020PaDiM}.
For IGD~\cite{chen2021unsupervised}, we pre-train the global and local models (see Figure~\ref{fig:framework}), where the patch position prediction loss in Eq.~\ref{eq:patch_prediction} is only fine-tuned for the local model.  
For PaDiM~\cite{defard2020PaDiM}, we pre-train the global model and use it to fine-tune the anomaly detection and segmentation models. 
For the training of IGD~\cite{chen2021unsupervised} and PaDiM~\cite{defard2020PaDiM}, we use the hyper-parameters suggested by the respective papers.
In our experiments, the local map for IGD is obtained by considering each 32$\times$32-pixel patch as an instance and apply our proposed self-supervised learning to it. 
The global map for IGD is computed based on the whole image sized as 256 $\times$ 256 pixels for Hyper-Kvasir, LAG and Covid-X datasets. For Liu et al.’s colonoscopy dataset, we only train the model globally with the image size 64 $\times$ 64. For the auto-encoder in IGD, we use the setup suggested in~\cite{chen2021unsupervised}, where the global model is trained with images of size 256$\times$256 pixels or 64 $\times$ 64 for Liu et al.’s colonoscopy dataset, and the local model is trained with image patches of size 32$\times$32. 
For PaDiM~\cite{defard2020PaDiM}, we only use the default setup in their work and compute the segmentation mask based on the images of size 256$\times$256 pixels for Hyper-Kvasir, LAG and Covid-X datasets, and 64 $\times$ 64 for Liu et al.’s colonoscopy dataset.


\subsection{Evaluation Measures}
\label{sec:evaluation_measures}

The anomaly detection performance is quantitatively assessed by the area under the receiver operating characteristic curve (AUROC), specificity, sensitivity and accuracy. 
AUROC assesses anomaly detection by varying the classification threshold and computing the area under the ROC curve. Sensitivity and specificity reflect the percentage of positives and negatives that are correctly detected. Accuracy shows the overall performance of correctly detected samples for both positive and negative images, where the classification threshold is estimated with a small validation set that contains 50 normal and 50 abnormal images that are randomly sampled from the testing set. Note that the validation set is only used for threshold estimation.
For anomaly segmentation, the performance is measured by Intersection over Union (IoU), Dice score, and Pro-score~\cite{bergmann2020uninformed}.  
IoU is computed by dividing the intersection by the union between the predicted segmentation and the ground truth mask. 
Dice also takes the predicted segmentation and the ground truth mask and divides two times their intersection by their sum.
Pro-score weights the ground-truth masks of different sizes equally~\cite{bergmann2020uninformed} to verify if both large and small abnormal lesions are accurately segmented. 

\begin{table}[t]
\centering
\resizebox{0.95\linewidth}{!}{
\begin{tabular}{@{}c c c c c c @{}}
\toprule \hline
Methods         & AUC & Specificity & Sensitivity  & Accuracy  \\ \hline\hline
DAE~\cite{masci2011stacked}             & 0.705  & 0.522 & 0.756  & 0.693   \\
OCGAN~\cite{perera2019ocgan}           & 0.813   & 0.691 & 0.811  & 0.795\\
f-AnoGAN~\cite{f-AnoGAN}        & 0.907     & 0.846  & 0.915   & 0.883  \\
ADGAN~\cite{liuphotoshopping}           & 0.913     & 0.879 & 0.946  & 0.893  \\
MS-SSIM~\cite{chen2021unsupervised}        & 0.917    & 0.857  & 0.925  & 0.912   \\
PANDA~\cite{reiss2021panda}   & 0.937   & 0.805  & 0.919 & 0.917 \\ 
CutPaste~\cite{li2021cutpaste} & 0.949   & 0.847  & 0.957 & 0.932 \\ 
\hline
PaDiM~\cite{defard2020PaDiM}   & 0.943   & 0.846 & 0.929 & 0.898 \\ 
CCD - PaDiM   & 0.978   & 0.923 & 0.961  & 0.967 \\
\textbf{PMSACL - PaDiM}   & \textbf{0.996}   & \textbf{0.966} & \textbf{0.981}  & \textbf{0.983} \\\hline
IGD~\cite{chen2021unsupervised}            & 0.939 & 0.858 & 0.913  & 0.906     \\ 
CCD - IGD      & 0.972   & 0.934 & 0.947  & 0.956  \\ 
\textbf{PMSACL - IGD}      & 0.995    & 0.947 &  0.965  & 0.972 \\
 \hline\bottomrule
\end{tabular}
}
\caption{\textbf{Anomaly detection} testing results on \textbf{Hyper-Kvasir} in terms of AUC, Specificity, Sensitivity and Accuracy. Best results are highlighted.}

\label{tab:detection_auc_HK}
\end{table}

\subsection{Anomaly Detection Results}
\label{sec:lesion_detection_results}

In this section, we show the anomaly detection results on all datasets.

\subsubsection{Hyper-Kvasir}

In Table~\ref{tab:detection_auc_HK}, we show the results of anomaly detection on Hyper-Kvasir dataset, where we present results from baseline UAD methods, including OCGAN~\cite{perera2019ocgan}, f-AnoGAN~\cite{f-AnoGAN}, ADGAN~\cite{liu2019photoshopping}, and deep autoencoder (DAE)~\cite{masci2011stacked} and its variant with MS-SSIM loss~\cite{chen2021unsupervised}. 
As discussed in Section~\ref{sec:anomaly_detection_Segmentation}, we choose IGD~\cite{chen2021unsupervised} and PaDiM~\cite{defard2020PaDiM} as the anomaly detector for evaluating our proposed PMSACL pre-training approach and compare it with our previously proposed CCD pre-training approach~\cite{tian2021constrained} to fine-tune IGD and PaDiM.

Comparing with the baseline UAD methods, the performance of PaDiM and IGD are improved using our PMSACL pre-trained encoder by around 5\% and 6\% AUC, which achieves SOTA anomaly detection AUC results of 99.6\% and 99.5\%, respectively, on Hyper-Kvasir. 
Comparing with our previously proposed CCD pre-training~\cite{tian2021constrained}, our proposed PMSACL pre-training improves the performance by 2.3\% and 1.8\% for PaDiM and IGD. 
This shows that our proposed MedMix and PMSACL loss improve the generalisation ability of the fine-tuning stage for anomaly detection and produce better constrained feature space of normal samples. 
Moreover, achieving SOTA results on two different types of anomaly detectors suggests that our self-supervised pre-training can produce good representations for both generative and predictive anomaly detectors.

OCGAN~\cite{perera2019ocgan} constrains the latent space based on two discriminators to force the latent representations of normal data to fall at a bounded area.  f-AnoGAN~\cite{f-AnoGAN} uses an encoder to extract the feature representations of a input image and use a GAN to reconstruct it. ADGAN~\cite{liuphotoshopping} uses two generators and two discriminators to produce realistic reconstruction of normal samples. 
These three methods achieve 81.3\%, 90.7\% and 91.3\% AUC on Hyper-Kvasir, respectively, which are well below our self-supervised PMSACL pre-training with IGD and PaDiM. 
Also, the recently proposed state-of-the-art (SOTA) methods PANDA~\cite{reiss2021panda} and CutPaste~\cite{li2021cutpaste} achieve significantly inferior performance than our PMSACL pre-trained anomaly detectors. Note that CutPaste uses a similar augmentation strategy as MedMix, but with inferior results, indicating the effectiveness of our proposed PMSACL self-supervised loss function.
Furthermore, PaDiM with our PMSACL pre-training can achieve the SOTA results of 96.6\% specificity, 98.1\% sensitivity and 98.3\% accuracy. 
It improves the previous PaDiM using CCD pre-training by 4.3\%, 2\% and 1.6\% for these three evaluation measures. 
Finally, PaDiM pre-trained with PMSACL significantly outperforms the PaDiM pre-trained with ImageNet~\cite{defard2020PaDiM} by 12\%, 5.2\% and 8.5\% in terms of these three evaluation measures.


\begin{table}[t]
\centering
\resizebox{0.9\linewidth}{!}{
\begin{tabular}{@{}c c c c c c c c@{}}
\toprule \hline
Methods         & AUC & Specificity & Sensitivity  & Accuracy  \\ \hline\hline
MS-SSIM~\cite{chen2021unsupervised}        & 0.823   & 0.257 & 0.937  & 0.774     \\
f-AnoGAN~\cite{f-AnoGAN}        & 0.778     & 0.565 & 0.899 & 0.763  \\
PANDA~\cite{reiss2021panda}   &  0.789   & 0.624    & 0.869  & 0.767 \\  
CutPaste~\cite{li2021cutpaste} & 0.745   & 0.372  & 0.788 & 0.685 \\ 
\hline
PaDiM~\cite{defard2020PaDiM}   & 0.688  & 0.314 & 0.809  &  0.673 \\ 
CCD - PaDiM      & 0.728   & 0.429 & 0.779  & 0.694  \\
\textbf{PMSACL - PaDiM}   & 0.761  &  0.466 & 0.877  & 0.753 \\\hline
IGD~\cite{chen2021unsupervised}            & 0.796  & 0.396 & 0.958  & 0.805     \\ 
CCD - IGD      & 0.874   & \textbf{0.572} & 0.944    &  0.875  \\ 
\textbf{PMSACL - IGD}      & \textbf{0.908}  & 0.531 & \textbf{0.979} & \textbf{0.884} \\ \hline\bottomrule
\end{tabular}
}
\caption{\textbf{Anomaly detection} testing results on \textbf{LAG} in terms of AUC, Specificity, Sensitivity, Precision and Recall. Best results are highlighted.}
\label{tab:detection_auc_LAG}
\end{table}

\subsubsection{LAG}

\begin{table}[t]
\centering
\resizebox{0.9\linewidth}{!}{
\begin{tabular}{@{}c c c c c c c @{}}
\toprule \hline
Methods         & AUC  & Specificity & Sensitivity  & Accuracy  \\ \hline\hline
DAE~\cite{masci2011stacked}             & 0.629*   & 0.733* & 0.554*  & 0.597*  \\
OCGAN~\cite{perera2019ocgan}           & 0.592*    & 0.716* &  0.534*  & 0.624*\\
ADGAN~\cite{liuphotoshopping}           & 0.730*     & 0.852* & 0.496*  & 0.713*  \\
f-AnoGAN~\cite{f-AnoGAN}        & 0.735      & 0.865  &  0.579  & 0.694 \\
PANDA~\cite{reiss2021panda}   & 0.719  & 0.846  & 0.551  & 0.671 \\  
CutPaste~\cite{li2021cutpaste} & 0.779   & 0.895  & 0.772 & 0.738 \\ 
\hline
PaDiM~\cite{defard2020PaDiM}   & 0.741  & 0.851 & 0.738  &  0.751 \\
CCD - PaDiM   & 0.789   & 0.946 & 0.792  & 0.767  \\
\textbf{PMSACL - PaDiM}   & 0.814  & 0.973 & 0.725  & 0.803  \\
\hline
IGD~\cite{chen2021unsupervised}            & 0.787  & 0.914 & 0.596  & 0.743   \\ 
CCD - IGD      &0.837  & 0.985 & 0.774  & 0.815  \\ 
\textbf{PMSACL - IGD}      & \textbf{0.851}  & \textbf{0.986} & \textbf{0.792} & \textbf{0.829}  \\ \hline\bottomrule
\end{tabular}
}
\caption{\textbf{Anomaly detection} testing results on \textbf{Liu et al.'s colonscopy} in terms of AUC, Specificity, Sensitivity and Accuracy.   * indicates that the model does not use ImageNet pre-training. Best results are highlighted.}
\label{tab:detection_auc_liu_etal}
\end{table}

We evaluate the performance of our PMSACL pre-training on the LAG dataset and show results on Table~\ref{tab:detection_auc_LAG}. 
Our PMSACL pre-training improves PaDiM and IGD AUCs by 7.3\% and 11.2\%, compared with their ImageNet pre-trained model, where the PMSACL pre-trained IGD achieves the SOTA results of 90.8\% AUC, 97.9\% sensitivity and 88.4\% accuracy.  
Comparing with our previous CCD pre-trained PaDiM and IGD~\cite{tian2021constrained}, our proposed PMSACL pre-trained PaDiM and IGD surpass them by 3.3\% and 3.4\% in terms of AUC. 
The MS-SSIM autoencoder~\cite{chen2021unsupervised}, f-AnoGAN~\cite{f-AnoGAN}, PANDA~\cite{reiss2021panda}, and CutPaste~\cite{li2021cutpaste} baselines achieve 82.3\%, 77.8\%,  78.9\%, and 74.5\% AUC, respectively, which are significantly inferior compared with our PMSACL pre-trained IGD. 
For LAG, IGD with both reconstruction and anomaly classification constraints can generally outperform PaDiM variants, indicating the superiority of IGD  when handling the subtle image features to detect glaucoma.

\subsubsection{Liu et al.'s Colonoscopy Dataset}

We further test our approach on Liu et al.'s colonoscopy dataset~\cite{liuphotoshopping}, as shown in Table~\ref{tab:detection_auc_liu_etal}. 
Our PMSACL pre-trained PaDiM improves the ImageNet pre-trained PaDiM by 7.3\% AUC, and CCD pre-trained PaDiM by 2.5\% of AUC. The IGD with the PMSACL pre-trained encoder achieves the SOTA result of 85.1\% AUC, surpassing the previous CCD and ImageNet pre-trained IGD by 1.4\% and 6.4\% AUC, respectively.

Compared with other UAD approaches, such as f-AnoGAN, ADGAN, OCGAN, PANDA, and CutPaste  that achieve 73.5\%, 73\%, 59.2\%, 70.2\% and 77.9\% AUC, our PMSACL pre-trained IGD and PaDiM produce substantially better results. 
The gap between PaDiM and IGD may be due to the low resolution of the images in this dataset, which hinders the PaDiM performance that requires dense intermediate feature maps. The additional results of the PMSACL pre-trained IGD are specificity of 98.6\%, sensitivity of 79.2\%, and accuracy of 82.9\%, which demonstrate the robustness of our proposed model.

\begin{table}[t]
\centering
\resizebox{0.85\linewidth}{!}{
\begin{tabular}{@{}ccccc@{}}
\toprule \hline
Methods         & AUC & Specificity & Sensitivity  & Accuracy   \\ \hline\hline
MS-SSIM~\cite{chen2021unsupervised}        & 0.634   & 0.572 & 0.406  & 0.577   \\
f-AnoGAN~\cite{f-AnoGAN}        & 0.669     & 0.718 & 0.365 & 0.532  \\
PANDA~\cite{reiss2021panda}   & 0.629   & 0.762  & 0.447  & 0.591 \\ 
CutPaste~\cite{li2021cutpaste} & 0.658   & 0.701  & 0.494 & 0.648 \\ 
\hline
PaDiM~\cite{defard2020PaDiM}  & 0.614 & 0.753 & 0.318 & 0.559    \\ 
CCD - PaDiM & 0.632 & 0.673 & 0.569 & 0.616   \\
\textbf{PMSACL - PaDiM}   & 0.658 & 0.749 & 0.467 & 0.615 \\
\hline
IGD~\cite{chen2021unsupervised}    & 0.699 & \textbf{0.885} & 0.490 & 0.688            \\ 
CCD - IGD      &  0.746 & 0.851 & 0.595 & 0.722  \\ 
\textbf{PMSACL - IGD}      & \textbf{0.872} & 0.863 & \textbf{0.775} & \textbf{0.813}   \\ \hline\bottomrule
\end{tabular}
}
\caption{\textbf{Anomaly detection} testing results on \textbf{Covid-X} in terms of AUC, Specificity, Sensitivity and Accuracy, respectively. Best results are highlighted.}
\label{tab:detection_auc_covid}
\end{table}

\subsubsection{Covid-X}

Table~\ref{tab:detection_auc_covid} shows that Covid-X results, where our PMSACL pre-trained PaDiM and IGD methods achieve 65.8\% and 87.2\% AUC on the Covid-X dataset, significantly surpassing their ImageNet pre-trained approaches by 4.4\% and 17.2\% AUC, and CCD pre-trained by 2.6\% and 12.6\% AUC.  
Moreover, our approaches achieve significantly better performance when compared against current SOTA MS-SSIM, f-AnoGAN, PANDA, and CutPaste baselines.
The small abnormal lesions in chest X-ray images are hard to detect, so the generative-based anomaly detector IGD can learn more effectively the fine-grained appearances of normal images, leading to better ability to detect unseen anomalous regions during testing with the SOTA results of 87.2\% AUC, 77.5\% of sensitivity and 81.3\% of accuracy. The PMSACL pre-trained IGD  achieves 86.3\% specificity, which is competitive with the result from IGD pre-trained with ImageNet. 
It can also be observed that our PMSACL pre-trained PaDiM and IGD improve sensitivity by 14.9\% and 28.5\%, when compared to the ImageNet pre-trained PaDiM and IGD.

\subsubsection{Variability in the Results}

We show on Table~\ref{tab:standard_deviation} the standard deviation computed from the AUC, specificity, sensitivity and accuracy results of five different trainings based on different model initialisation of the PMSACL pre-trained PaDiM detector. These results in Table~\ref{tab:standard_deviation} should be studied together with the Tables~\ref{tab:detection_auc_HK},~\ref{tab:detection_auc_LAG},~\ref{tab:detection_auc_covid}. In general, we conclude that the differences between the methods described in the sections above can be considered significant in most cases given that 
the standard deviation only varies from $0.5\%$ to $1.5\%$.

\begin{table}[t]
\centering
\resizebox{0.9\linewidth}{!}{
\begin{tabular}{@{}c c c c c c c c@{}}
\toprule \hline
Dataset         & AUC & Specificity & Sensitivity  & Accuracy  \\ \hline\hline
Hyper-Kvasir~\cite{borgli2020hyperkvasir}  &   0.0084   & 0.0079   & 0.0127 & 0.0036    \\
LAG~\cite{li2019attention}        & 0.0163    & 0.0085 &  0.0105 & 0.0121  \\
Covid-X~\cite{wang2020covid}    &  0.0107  & 0.0149 & 0.0092 & 0.0171 \\ 
 \hline\bottomrule
\end{tabular}
}
\caption{The standard deviation of five-run experimental results on the Hyper-Kvasir, LAG and Covid-X based on the PMSACL pre-trained PaDiM anomaly detector. This results should be studied together with the results shown in Tables~\ref{tab:detection_auc_HK},~\ref{tab:detection_auc_LAG},~\ref{tab:detection_auc_covid}.
}
\label{tab:standard_deviation}
\end{table}

\begin{figure}[t]
\begin{center}
\includegraphics[width=0.95 \linewidth]{ 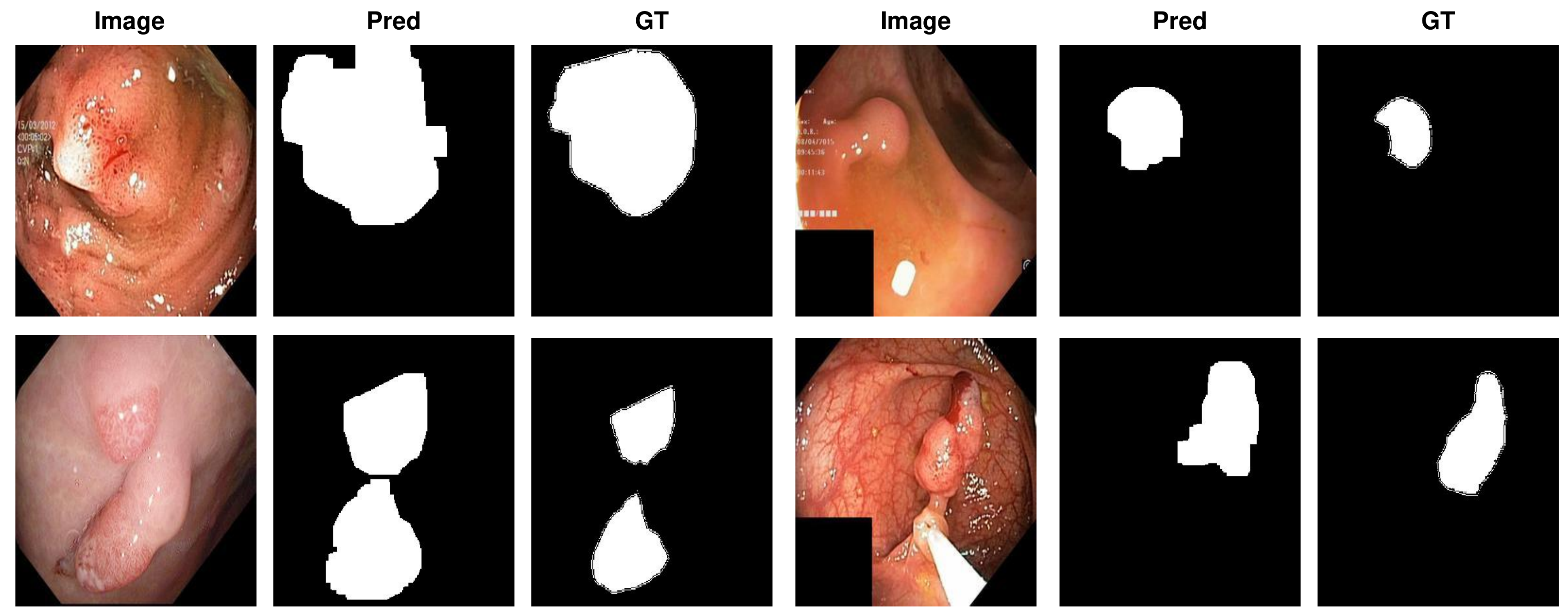}
\end{center}
   \caption{Localisation of four abnormal images from Hyper Kvasir~\cite{li2019attention}, with their predictions (Pred) and ground truth annotations (GT), using PaDiM  with PMSACL pre-training.}
\label{fig:heatmap_hk}
\end{figure}

\begin{figure}[t]
\begin{center}
\includegraphics[width=0.95 \linewidth]{ 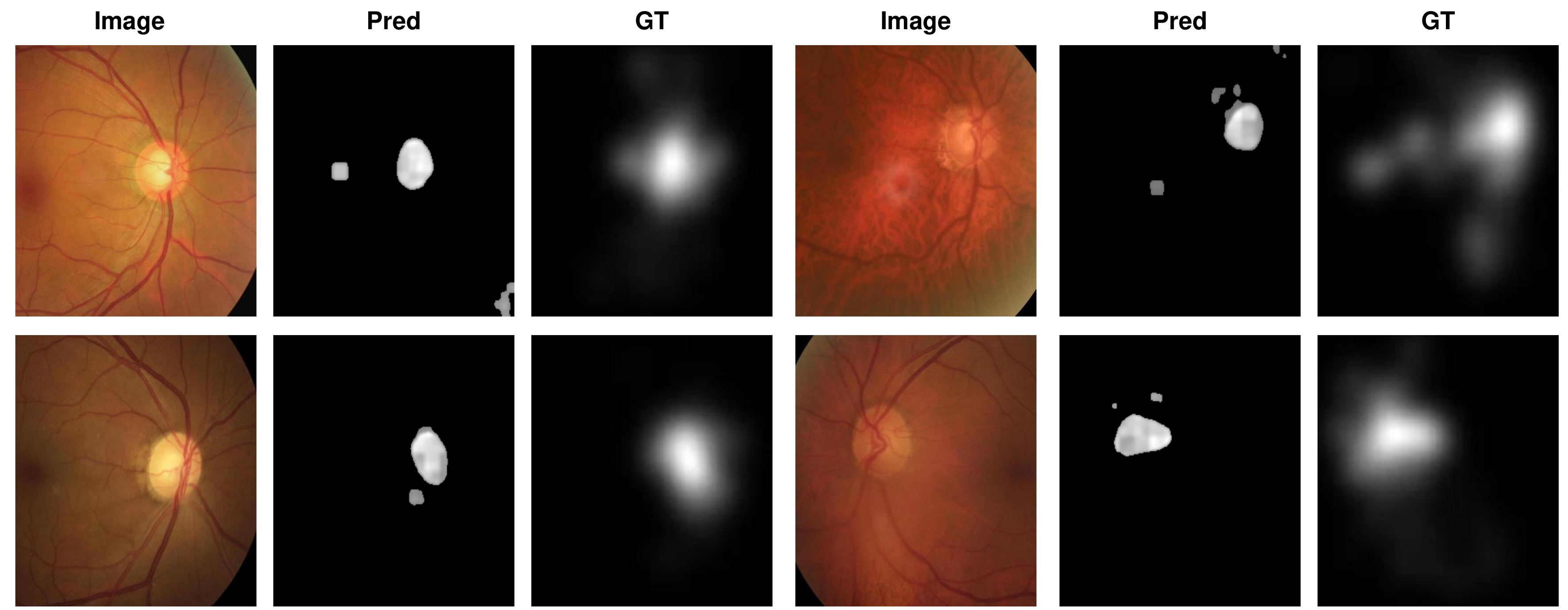}
\end{center}
   \caption{
  Localisation of four abnormal images from LAG~\cite{li2019attention}, with their predictions (Pred) and ground truth attention maps (GT), using IGD with PMSACL pre-training. }
\label{fig:heatmap_lag}
\end{figure}

\subsection{Anomaly Localisation Results}

In this section, we show the anomaly localisation results on Hyper-Kvasir and LAG.


\subsubsection{Hyper-Kvasir}

We demonstrate the anomaly localisation performance on Hyper-Kvasir on Table~\ref{tab:localisation_auc_HK}. Following~\cite{tian2021constrained}, we randomly sample 100 abnormal images from the test set and compute the mean segmentation performance over five different such groups of 100 images.
The proposed PMSACL pre-training improves the IGD and PaDiM by 1.2\% and 2.8\% IoU compared with the CCD pre-training, and 8.1\% and 6.4\% IoU with respect to the ImageNet pre-training, respectively. 
In addition, our PMSACL pre-trained PaDiM shows the SOTA result of 40.6\% IoU and 55.4\% Dice, demonstrating the effectiveness of our PMSACL approach for abnormal lesion segmentation. The CCD version of PaDiM achieves the SOTA result of 88.1\% Pro-score.

\begin{table}[t]
\centering
\resizebox{0.75\linewidth}{!}{
\begin{tabular}{@{}cccc@{}}
\toprule \hline
 Methods & IoU & Dice & Pro  \\ \hline \hline
 PaDiM~\cite{defard2020PaDiM}         &  0.341 &  0.475   & 0.803 \\
 CCD - PaDiM       & 0.378   & 0.497 &\textbf{0.881}  \\
\textbf{PMSACL - PaDiM}       & \textbf{0.406} & \textbf{0.554}  & 0.854 \\ \hline
IGD~\cite{chen2021unsupervised}       &  0.303  & 0.417 & 0.794  \\ 
CCD - IGD       &  0.372  & 0.502  & 0.865 \\ 
\textbf{PMSACL - IGD}       & 0.384  & 0.521  & 0.876  \\ 
\hline 
 \bottomrule \hline
\end{tabular}%
}
\caption{\textbf{Anomaly localisation:} Mean IoU, Dice and PRO-AUC testing results on \textbf{Hyper-Kvasir} on 5 different groups of 100 images with ground truth masks.Best results for each case are highlighted.}
\label{tab:localisation_auc_HK}
\end{table}

\subsubsection{LAG}

We further demonstrate the segmentation results on LAG dataset on Table~\ref{tab:localisation_auc_LAG}. The PMSACL pre-trained IGD achieves the SOTA result of 51.6\% IoU, 66.7\% Dice and 69.3\% Pro-score, showing that our model can effectively segment different types of lesions, such as colon polyps or optic disk and cup with Glaucoma. Moreover, PaDiM pre-trained with PMSACL improves PaDiM pre-trained with CCD and ImageNet by 1.3\% and 4.8\% IoU, respectively. Also, PaDiM with PMSACL pre-training achieves 64.3\% Dice and 62.8\% Pro-score, which are comparable to the SOTA results by the PMSACL pre-trained IGD. 
\begin{table}[t]
\centering
\resizebox{0.75\linewidth}{!}{
\begin{tabular}{@{}cccc@{}}
\toprule \hline
 Methods & IoU & Dice & Pro  \\ \hline \hline
PaDiM~\cite{defard2020PaDiM}        & 0.427 & 0.579  & 0.596\\
CCD - PaDiM       & 0.462   & 0.612   & 0.634 \\
\textbf{PMSACL - PaDiM}       & 0.475  & 0.643   & 0.628 \\\hline  
IGD~\cite{chen2021unsupervised}        & 0.409 &  0.539 & 0.603 \\
CCD - IGD       & 0.509    & 0.645 &  0.677 \\ 
\textbf{PMSACL - IGD}      &  \textbf{0.516} &\textbf{0.667}  &  \textbf{0.693} \\ 
\hline 
 \bottomrule \hline
\end{tabular}%
}
\caption{\textbf{Anomaly localisation:} Mean IoU, Dice and Pro-AUC testing results on abnormal samples from \textbf{LAG} test set. Best results are highlighted.}
\label{tab:localisation_auc_LAG}
\end{table}

\subsection{Qualitative Results}

In this section, we show examples of anomaly localisation and detection results, and t-SNE results displaying the distribution of image representations of the normal and pseudo abnormal classes in the feature space.

\subsubsection{Anomaly Localisation and Detection Visual Results}

The visualisation of polyp localisation results of PaDiM with PMSACL pre-training on Hyper-Kvasir~\cite{borgli2020hyperkvasir} is shown in Fig.~\ref{fig:heatmap_hk}. 
Notice that our model can effectively localise colon polyps with various sizes and shapes. 
We also show the localisation results based on the pixel-level anomaly scores of IGD with PMSACL pre-training on the LAG dataset in Fig.~\ref{fig:heatmap_lag}.

\begin{figure*}[t]
\begin{center}
\includegraphics[width=0.9\linewidth]{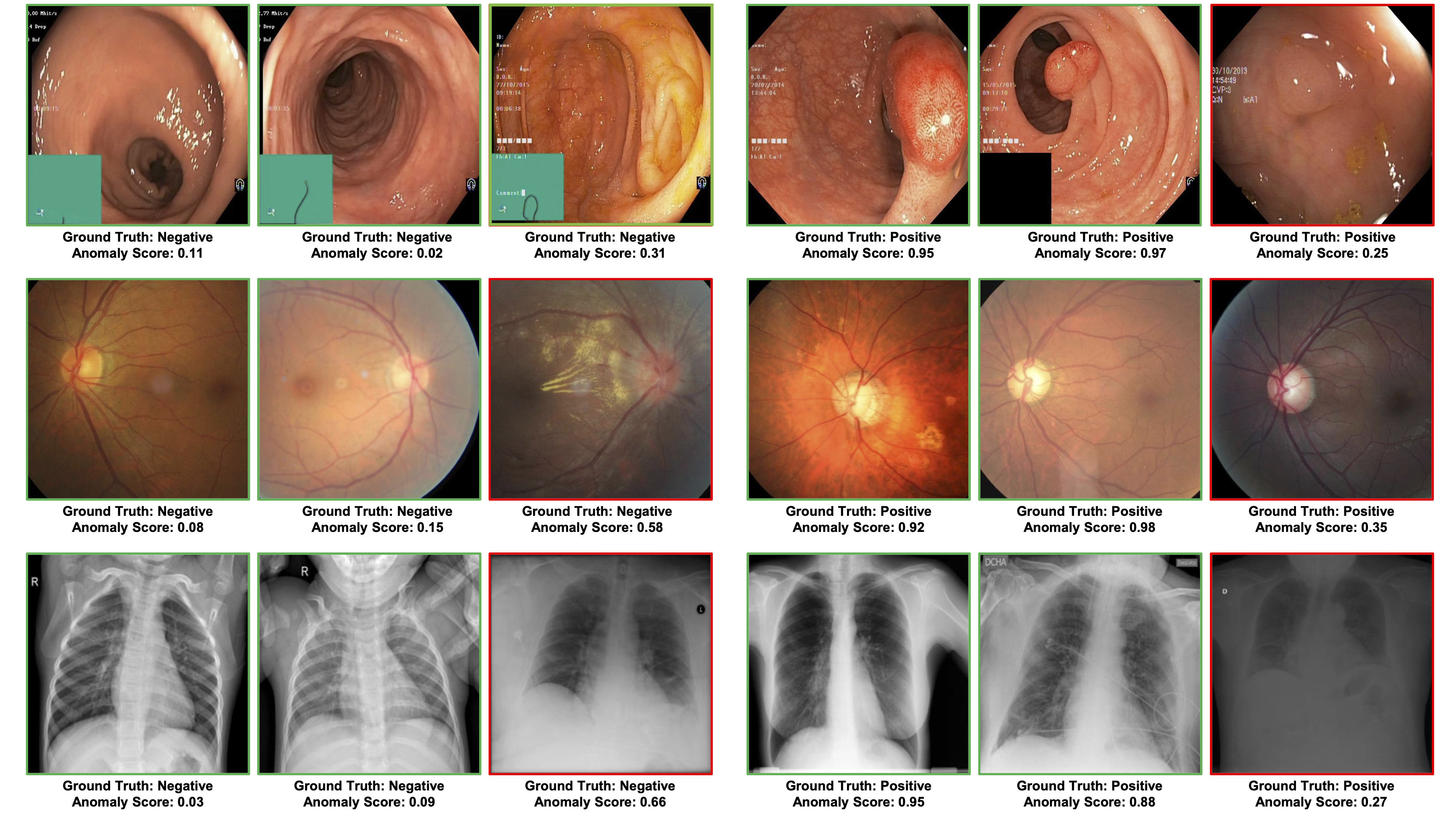}
\end{center}
   \caption{Visual detection results and anomaly scores produced by the PMSACL pre-trained IGD on three different datasets: Hyper-Kvasir (top), LAG (middle), Covid-X (bottom). 
   Anomaly scores $> 0.5$ classifies the image as positive, otherwise, the image is classified as negative. 
   Correctly classified images are marked with green boxes, and incorrectly classified cases are marked with red boxes.}
\label{fig:prediction_scores}
\end{figure*}

The visual anomaly detection results of IGD pre-trained with PMSACL on the Hyper-Kvasir~\cite{borgli2020hyperkvasir} test set is shown in Figure~\ref{fig:prediction_scores}.

\subsection{Ablation Study}

In this section, we study the roles played by PMSACL components. 
We start by investigating the loss terms in~\eqref{eq:full_loss}. 
Then we study the impact of using different types of strong data augmentation to generate the pseudo abnormal images and the number of pseudo abnormal classes in MedMix (i.e., the size $|\mathcal{A}|$ in~\eqref{eq:full_loss}). We also compare our approach with other recently proposed self-supervised pre-training approaches.


\begin{table}[h]
\centering
\resizebox{1.0\linewidth}{!}{
\begin{tabular}{ccccc|cc}
\toprule\hline
CCD & MedMix & $\ell_{PMSACL}$  & $\ell_{ctr}$  & $\kappa(n,m)$    & AUC - Hyper  & AUC - LAG \\ \hline \hline
\checkmark       &           &        &                 &       &   0.978   &    0.728     \\  
\checkmark       &  \checkmark         &          &                  &    &   0.985   &    0.739     \\
\checkmark       &    \checkmark              &  \checkmark      &              &    &   0.990  &    0.745     \\
\checkmark    &  \checkmark   & \checkmark           &  \checkmark                   &   &   0.993      &  0.753     \\
\checkmark    &  \checkmark       &\checkmark   & \checkmark   &  \checkmark   &  0.996  & 0.761 \\ \hline \bottomrule
\end{tabular}
}
\caption{\textbf{Ablation study of the PMSACL loss terms} on the test sets of Hyper-Kvasir and LAG, using PaDiM~\cite{defard2020PaDiM} as anomaly detector, with our MedMix as strong augmentations. Please note that the CCD~\cite{tian2021constrained} consists of the patch position loss $\ell_{pos}$, augmentation loss $\ell_{aug}$, and the standard contrastive loss, with \textbf{MedMix} as the strong augmentations. }
\label{tab:ablation}
\end{table}
\subsubsection{Loss Terms in PMSACL pre-training} 


We present an ablation study that shows the influence of each term of our proposed PMSACL pre-training in~\eqref{eq:full_loss} following PaDiM fine-tuning in Table~\ref{tab:ablation} on Hyper-Kvasir and LAG datasets.  
Starting from our previous CCD framework~\cite{tian2021constrained}, which includes the CCD contrastive loss, the strong augmentation loss $\ell_{aug}$, and the patch location prediction loss $\ell_{pos}$, the performance can reach 97.8\% and 72.8\% AUC on Hyper-Kvasir and LAG datasets, respectively. Moreover, we notice that MedMix can improve the AUC on both datasets by 1\%. Our proposed $\ell_{PMSACL}$ without temperature scaling strategy $\kappa(n,m)$ and multi-centring loss $\ell_{ctr}$ can improve 0.5\% and 0.6\% of AUC on both datasets. Then, adding the proposed multi-centring loss $\ell_{ctr}$ and temperature scaling $\kappa(n,m)$ from~\eqref{eq:temperature_calibration} further boosts the performance to the state-of-the-art results of 99.6\% and 76.1\% AUC on both datasets. This indicates that the joint training of $\ell_{PMSACL}$ and $\ell_{ctr}$ with temperature scaling strategy can learn better fine-grained low-dimensional features for the downstream anomaly detectors (i.e., producing denser cluster for normal images).


\begin{figure}[t]
 \centering    
 \includegraphics[width=.65\linewidth]{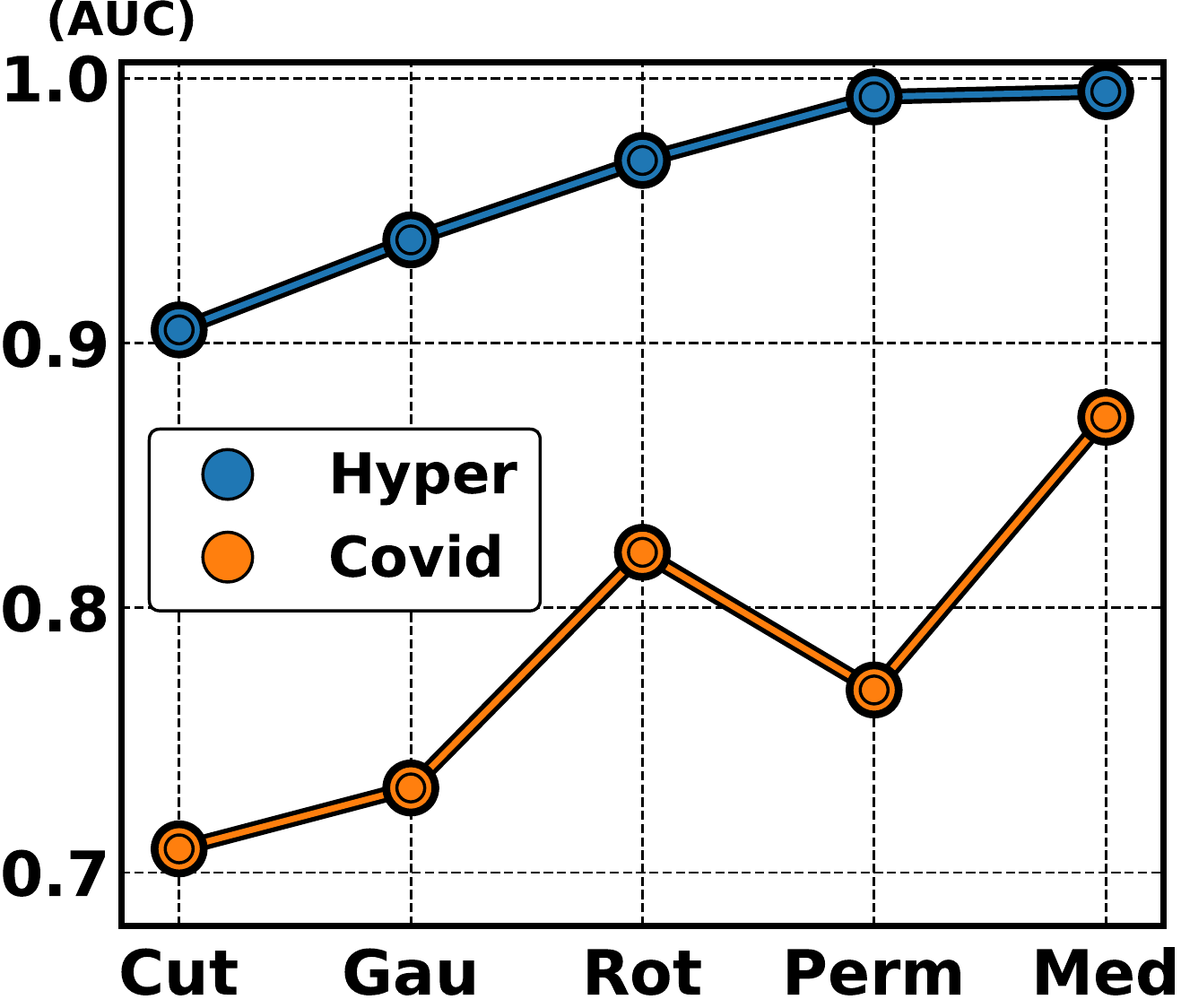}
 \vspace{-10pt}
\caption{Anomaly detection testing results in terms of different types of strong augmentations (i.e., Cutmix, Gaussian noise, Rotation, Permutation, and our MedMix) on Hyper-Kvasir and Covid-X, where
our PMSACL is used as self-supervised pre-training, and
IGD~\cite{chen2021unsupervised} is used as the anomaly detector.}
    
\label{fig:ablation_transformation}
\end{figure}
\begin{figure}[t]
 \centering    
 \includegraphics[width=.75\linewidth]{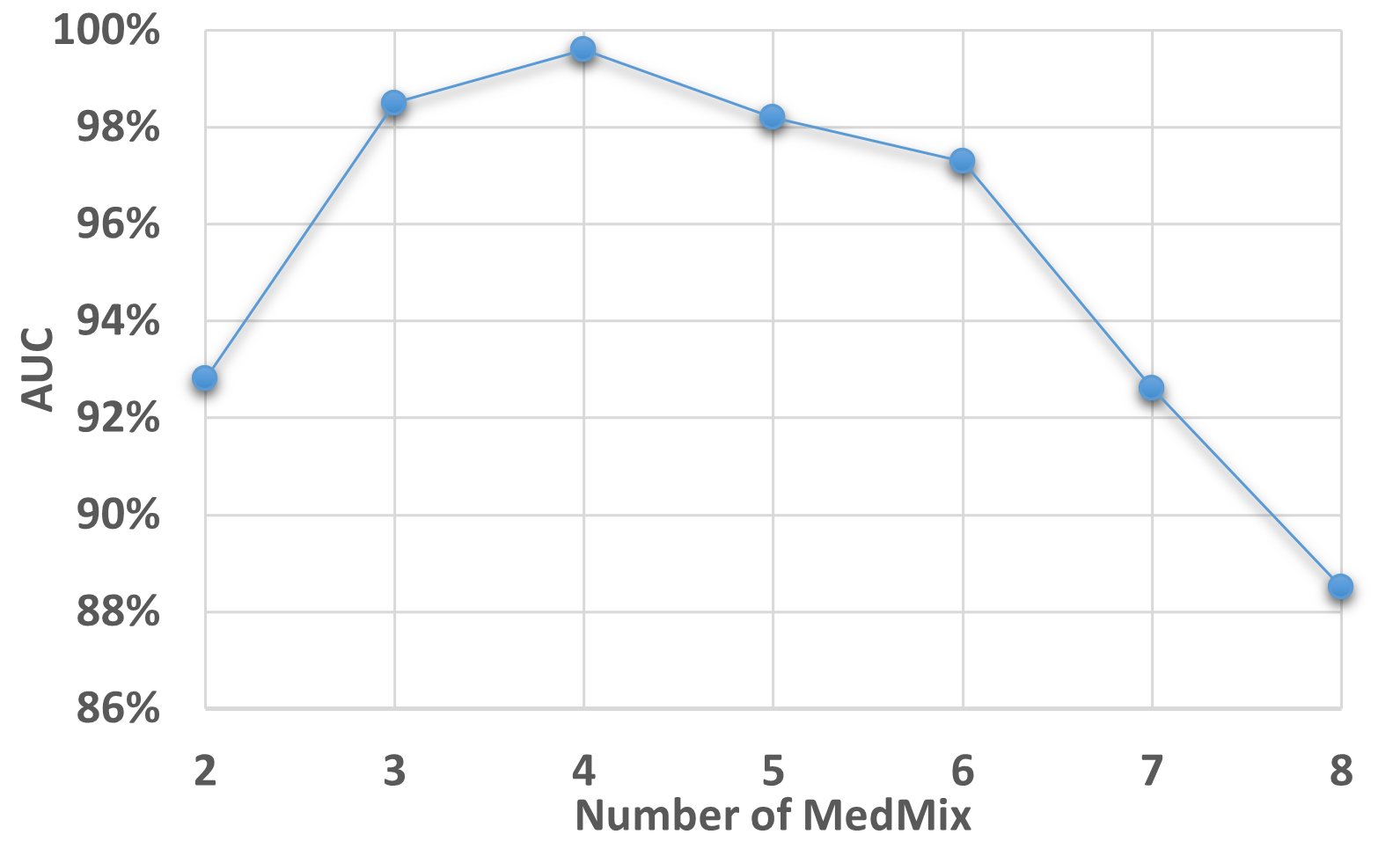}
 \vspace{-10pt}
\caption{Influence of the number of MedMix augmentation distributions $|\mathcal{A}|$ in~\eqref{eq:full_loss} on the AUC results of Hyper-Kvasir testing set, where PaDiM~\cite{defard2020PaDiM} is used as the anomaly detector. 
}
   \label{fig:ablation_num_aug}
\end{figure}

\begin{table}[t]
\centering
\resizebox{0.95\linewidth}{!}{
\begin{tabular}{@{}c c c c c c c @{}}
\toprule \hline
Methods         & AUC  & Specificity & Sensitivity  & Accuracy  \\ \hline\hline
ImageNet           & 0.943   & 0.846 & 0.929 & 0.898  \\
SimCLR~\cite{simclr_paper}          & 0.945  & 0.794 & 0.942   & 0.914 \\
Rot-Net~\cite{golan2018deep}          & 0.938  & 0.856 & 0.905  & 0.905  \\
CSI~\cite{tack2020csi}   & 0.946   & 0.952   & 0.914  & 0.933  \\
DROC~\cite{sohn2020learning} & 0.931   & 0.954 & 0.881   & 0.914 \\
SupCon~\cite{khosla2020supervised}   & 0.946   & 0.912 & 0.953  & 0.928  \\
CCD~\cite{tian2021constrained}          & 0.978   & 0.923 & 0.961  & 0.967 \\
\hline
PMSACL  & \textbf{0.996}   & \textbf{0.966} & \textbf{0.981}  & \textbf{0.983}   \\ \hline\bottomrule
\end{tabular}
}
\caption{Ablation studies with different self-supervised pre-training approaches on Hyper-Kvasir testing set. PaDiM~\cite{defard2020PaDiM} is used as the anomaly detector. Best results are highlighted.}
\label{tab:ablation_ssl}
\end{table}

\subsubsection{Strong Augmentations}

In Fig.~\ref{fig:ablation_transformation}, we explore the influence of strong augmentation strategies, represented by rotation, permutation, cutout, Gaussian noise and our proposed MedMix on the AUC results of Hyper-Kvasir and Covid-X datasets, based on our self-supervised PMSACL pre-training with IGD as anomaly detector. The performance of our MedMix reaches the SOTA results of 99.5\% and 87.2\% on those datasets.
The second best AUC (96.9\%) on Hyper-Kvasir uses random permutations, which were used in CCD pre-training~\cite{tian2021constrained}, producing an AUC 0.2\% worse than our MedMix. For Covid-X, rotation is the second best data augmentation approach with an AUC result that is 5.1\% worse than MedMix. 
Other approaches do not work well with the appearance characteristics of X-ray images, yielding significantly worse results than our MedMix on Covid-X. 
These results suggest that the use of MedMix as the strong augmentation yields the best AUC results on different medical image benchmarks.

\subsubsection{MedMix Augmentations}

In Fig.~\ref{fig:ablation_num_aug}, we explore the influence of the number of MedMix augmentation distributions (i.e., $|\mathcal{A}|$) on the AUC results of Hyper-Kvasir, based on our self-supervised PMSACL pre-training and PaDiM anomaly detector. Our model achieves the best performance when $|\mathcal{A}|=4$ strong augmentation distributions, when it reaches around 98\% to 99\% AUC. The AUC declines when $|\mathcal{A}|>5$ or $|\mathcal{A}|<3$. The performance deterioration when $|\mathcal{A}|<3$ is due to an insufficient number of pseudo abnormal training regions from the strong augmentation distributions. When the number of strong augmentation distributions increases to $|\mathcal{A}|>5$, the pseudo abnormalities may hide most of the normal image regions, causing the model to become over-confident when classifying the pseudo abnormal regions.

\subsubsection{Other Self-supervised Methods}

In Table~\ref{tab:ablation_ssl}, we show the results of different pre-training approaches with PaDiM as anomaly detector, on Hyper-Kvasir testing set. 
It can be observed that our PMSACL approach surpasses the previous SOTA CCD pre-training~\cite{tian2021constrained} by 2.2\% AUC. Other pre-training methods proposed in computer vision (e.g., ImageNet pre-training, SimCLR~\cite{simclr_paper}, Rot-Net~\cite{golan2018deep}) achieve worse results than CCD and PMSACL. An interesting point in this comparison is the relatively poor result from ImageNet pre-training, suggesting that it may not generalise well for anomaly detection in medical images. Finally, our PMSACL achieves better results than previous SOTA UAD SSL approaches CSI~\cite{tack2020csi} and DROC~\cite{sohn2020learning} by about 4\% to 5\% AUC, indicating the effectiveness of our new contrastive loss. We also compare the SOTA supervised contrastive learning  SupCon~\cite{khosla2020supervised}, which re-formulates the contrastive loss as a supervised task. To validate the effectiveness of our proposed PMSACL contrastive loss, we adapted SupCon to our pseudo multi-class pre-training paradigm for performance comparison. The anomaly detection performance of our PMSACL significantly surpasses SupCon. We argue that such performance improvement is due to the fact that SupCon does not contrast the samples from the same classes, missing the chance of learning fine-grained normality features between those samples.   

\begin{table}[h]
\centering
\resizebox{0.9\linewidth}{!}{
\begin{tabular}{@{}c|c|c|c|c@{}}
\toprule \hline
Methods         & Hyper  & Liu et al. & LAG  & Covid \\ \hline\hline
Random Centre           & 0.902 & 0.792 & 0.803  & 0.795     \\ 
Equidistant Centre      & 0.912   & 0.772 & 0.825  & 0.814  \\ 
Re-estimate Centre     & 0.985  & \textbf{0.864} & 0.878  & \textbf{0.876 }\\ \hline
\textbf{Centres from untrained encoder}      & \textbf{0.995}    &\textbf{0.851} &  \textbf{0.908}  & \textbf{0.872} \\
 \hline \bottomrule
\end{tabular}
}
\caption{\textbf{Anomaly detection:} \textbf{AUC} test results that compares different methods for defining the class centres, using \textbf{IGD} as anomaly detector, on Hyper-Kvasir, Liu et al.'s colonoscopy, LAG, and Covid-X.  }
\label{tab:centers_difference}
\end{table}

\subsubsection{Centering Strategies}
As aforementioned, UAD models often suffer from the issue of catastrophic
collapse~\cite{ruff2018deep,chen2021unsupervised}, where all training samples are projected to a single point in the representation space. To avoid such an issue, previous UAD methods used a pre-defined class centre and kept it fixed throughout the entire training.  We also observe such catastrophic
collapse issue when iteratively updating the class centres during our experiments. In particular, the optimisation processes often fail with some of the random seeds and we observed that roughly only one out of five optimisation processes can be successful.  
Another major drawback of iteratively updating the latent centre through the training process is that such a process often requires a significant amount of computation time due to the large number and resolution of training medical images. From our experiment of re-estimating centres, the overall training time increased about three times. 
We found that pre-defining the class centres enables the best performance.
As shown in Table~\ref{tab:centers_difference}, we added a new ablation study to compare the performance with regard to randomly defined centres, pre-defined equidistant vectors as centres, re-estimated centres over multiple epochs, and our approach of pre-defined centres from an untrained model, on all four medical anomaly detection datasets.  The inferior performance of randomly defined centres and pre-defined equidistant centres is because those centres may fall in sub-spaces that are too distant from the actual representation distribution from the beginning of the training process, leading to ineffective optimisation and challenging the optimisation process. Lastly, re-estimating centres every 5 epochs achieves comparable performance than our approaches but sacrificing significant amount of training efficiency and stability. 

\section{Conclusion}

In this paper, we proposed a new self-supervised pre-training approach, namely PMSACL, for UAD methods applied to MIA problems.
PMSACL is based on a new contrastive learning optimisation to learn multiple classes of normal and pseudo abnormal images, formed with the proposed MedMix data augmentation that simulates medical abnormalities.
After pre-training a UAD model using our PMSACL, we fine-tune it with two SOTA anomaly detecting approaches.
Experimental results indicate that our PMSACL pre-training can effectively improve the performance of anomaly detection and segmentation on several medical datasets for both anomaly detectors.
In the future, we plan to design a new anomaly detector that suits better the characteristics of our self-supervised PMSACL pre-training. 

\clearpage
{\small
\bibliographystyle{ieee_fullname}
\bibliography{egbib}
}

\end{document}